\documentclass[]{pasj02} 
\usepackage{fix-cm}
\usepackage[switch,mathlines]{lineno}
\usepackage{url}

\jyear{2026}
\Received{}%
\Accepted{}%

\begin{document} 

\title{State-Space Modeling of X-Ray Spectral Component Variations in a Black Hole Binary}

\author{
 Tomoki \textsc{Omama}
 \altaffilmark{1}\altemailmark
\orcid{0000-0002-4192-4424}
 \email{omama.tomoki@gmail.com}
}
\author{Masahiro~\textsc{Tsujimoto}
\altaffilmark{2,3}\altemailmark
\orcid{0000-0002-9184-5556} 
\email{tsujimoto.masahiro@jaxa.jp} 
}
\author{Makoto~\textsc{Uemura}
\altaffilmark{1}\altemailmark
\orcid{0000-0002-7375-7405}
\email{uemuram@hiroshima-u.ac.jp}
}
\altaffiltext{1}{Hiroshima Astrophysical Science Center, Hiroshima University, 1-3-1 Kagamiyama, Higashi-Hiroshima 739-8526, Japan}
\altaffiltext{2}{Institute of Space and Astronautical Science, Japan Aerospace Exploration Agency, 3-1-1 Yoshinodai, Chuo-ku, Sagamihara, Kanagawa 252-5210, Japan}
\altaffiltext{3}{Department of Astronomy, Graduate School of Science, The University of Tokyo, Bunkyo-ku, Tokyo 113-0033, Japan}


\maketitle

\begin{abstract}
 Observed multi-band X-ray light curves of celestial objects are often mixtures of
 multiple spectral components with different physical origins. This spectral mixture has
 been a major challenge in conventional timing analysis. We propose a framework for
 joint spectral and timing modeling, which integrates spectral constraints into
 a dynamic factor model within the state-space framework. Instead of analyzing the
 observed multi-band light curves, this framework allows us to evaluate the time- and
 frequency-domain behaviors of the latent variables that represent the assumed physical
 spectral components. 
 Applying this framework to the Neutron Star Interior Composition Explorer
 (NICER) data of the black hole binary MAXI J1820$+$070 in the low/hard state, we
 estimate the variability of three modeled components: Comptonization, disk blackbody,
 and soft excess. Based on this decomposition, we identify a signal propagation in the
 sequence of disk $\rightarrow$ Comptonization $\rightarrow$ soft excess, with discrete
 time lags of 0.1~s and 0.2~s relative to the disk. These timescales consistently point
 to a truncation radius of $\sim 100 r_{\rm g}$ of the accretion disk, which is in
 agreement with some previous studies. This framework is widely applicable to data taken
 with photon-counting detectors, offering a robust approach to characterize the
 variability of physical components.
\end{abstract}

\section{Introduction}
\label{section:intro}

Variability in emission provides crucial insights into astrophysical phenomena. One
example is black hole binaries (BHBs), in which rapid X-ray variability is used to infer
the geometry and dynamics of accretion flows near compact objects. Traditionally, light
curves are constructed in multiple energy bands, and their temporal characteristics are
analyzed \citep{vanderklisRapidXrayVariability2006}. However, the emission in each energy band is often a mixture of several
spectral components with different physical origins. These spectral components are
typically analyzed using time-sliced spectroscopy, which often sacrifices temporal
resolution to accumulate a sufficient number of photons. We need a method to
characterize the rapid variability of individual spectral components, rather than that
of individual energy bands, in order to better constrain the underlying physical
processes. The purpose of this paper is to present a general framework for such an
analysis, applicable to data obtained with photon-counting detectors that have
spectroscopic capability, such as those commonly used in high-energy astrophysics.

We treat data as a two-dimensional histogram of time and energy, hereafter referred to
as an ``energy spectrogram'', analogous to a conventional audio spectrogram. To extract
physical insights, time-series statistical modeling approaches offer a powerful
framework. Among them, we use the state-space model, in which the time series of
observed variables are modeled separately from hidden variables called ``latent
variables''. This model has been applied in a wide range of fields, including control
engineering, signal processing, economics, aerospace engineering, and meteorology and
Earth sciences. 
However, the application in astrophysics is limited to a few literature
such as continuous autoregressive models for identifying representative timescales of
variability \citep{kellySTOCHASTICMODELLUMINOSITY2011} and multi-band modeling for
estimating time delays \citep{huModelingStochasticVariability2020}. These pioneering
studies characterize only the collective or empirical behavior of multi-band
observations and do not isolate the underlying physical spectral components.

The state-space model is highly flexible for multivariate time-series analysis
\citep{kitagawaIntroductionTimeSeries2022}. Among the various modeling formulations
within this framework, the dynamic factor model
\citep{penaIdentifyingSimplifyingStructure1987a,banburaMaximumLikelihoodEstimation2014}
is particularly powerful, as it is capable of representing a large number of observed
variables through a small number of underlying, unobserved latent variables. Using this
model, we can describe the intrinsic variations by associating the assumed physical components 
with the latent variables and multi-band light curves with the observed variables.

We demonstrate the effectiveness of this model by applying it to the actual data of
a BHB MAXI J1820$+$070 obtained with the Neutron Star Interior Composition Explorer
(NICER) during its low/hard state.

\medskip

The remainder of this paper is as follows. In Section~\ref{section:methodology}, we
present the framework in more detail with the mathematical formulation of the energy
spectrogram (\S~\ref{subsection:formulation_of_energy_spectrogram}) and the
specification of the state-space model (\S~\ref{subsection:state-space_model}).
Section~\ref{section:observations} describes the data
(\S~\ref{subsection:nicer_observations_and_data_reduction}) and the time-averaged
spectral analysis (\S~\ref{subsection:spectral_analysis}).
Section~\ref{section:model_configuration} specifies the model configuration, including
the energy band selection, the structured observation and system equations, the choice
of prior distributions, and the sampling setup. In Section~\ref{section:results}, we
present the reconstructed light curves of the assumed physical components, followed by
the characterization of their frequency-domain and time-domain dynamics, and the
reproduction of their time-averaged energy spectra. Section~\ref{section:discussion}
provides the physical insights obtained with this method.
Finally, a summary of this work is given in Section~\ref{section:summary_and_outlook}.

\section{Methodology}
\label{section:methodology}

\subsection{Formulation of Energy Spectrogram}
\label{subsection:formulation_of_energy_spectrogram}

In data taken with a photon-counting device, individual photons carry both arrival time
and energy information. We can construct a two-dimensional histogram in the time and
energy domains, which we call an ``energy spectrogram''. We denote the observed count
in a cell as $c_{nm}$, 
where the total observation duration is segmented into $n = 1,
\dots, N$ time bins, and the energy range is divided into $m = 1, \dots, M$ energy
bands. The energy spectrogram $C$ can be represented as an $M \times N$ matrix as
\begin{equation}
\label{e01}
  C = 
  \begin{pmatrix}
    c_{11} & c_{21} & \cdots & c_{N1} \\
    c_{12} & c_{22} & \cdots & c_{N2} \\
    \vdots & \vdots & \ddots & \vdots \\
    c_{1M} & c_{2M} & \cdots & c_{NM}
  \end{pmatrix}.
\end{equation}

As common in the statistical modeling framework, we treat the observed counts as
realizations of random variables and infer their joint probability distribution $p(C)$.
Let the $n$-th column of the matrix $C$ be an $M$-dimensional observation vector $y_n =
[c_{n1}, c_{n2}, \dots, c_{nM}]^T$, which represents the multi-band light curve at time
bin $n$. The joint probability distribution of the entire energy spectrogram $p(C)$ is
equivalent to the joint probability distribution of these multi-band light curves:
\begin{equation}
  p(C) = p(y_1, y_2, \dots, y_N).
\end{equation}
Because this formulation retains and simultaneously models both the temporal evolution
(timing) and the energy distribution (spectroscopy), we refer to the estimation of
$p(y_1, \dots, y_N)$ as ``joint spectral and timing modeling.''

\subsection{State-Space Model}
\label{subsection:state-space_model}

The state-space model consists of two relations: one is the system model, which dictates
the temporal evolution of the latent variables, and the other is the observation model,
which defines the mapping from the latent variables to the observed variables. Two
major assumptions are often adopted: the linearity of the relations and the Gaussianity
of the underlying noise, which is classified as the Linear Gaussian State-Space Model
(LGSSM). The LGSSM allows for highly efficient filtering, smoothing, and exact
likelihood calculation via the K\'{a}lman filter algorithm.

\subsubsection{Linear Gaussian State-Space Model}
\label{subsubsection:linear_gaussian_state_space_model}

Let the latent variable $x_n$ be a $k$-dimensional vector representing the physical
states, and the observation variable $y_n$ be a $\ell$-dimensional vector representing the
multi-band light curve at the time bin $n$. The formulation of the LGSSM is expressed as
follows:
\begin{equation}
  \begin{split}
  x_n &= F x_{n-1} + v_n, \\
  y_n &= H x_n + w_n,
  \end{split}
  \label{eq:linear_gaussian_state_space_model}
\end{equation}
where $F$ is the $k \times k$ matrix for the system model, and $H$ is the $\ell \times k$
matrix for the observation model. Each relation comes with a noise term ---the system
noise $v_n$ and the observation noise $w_n$---, which are assumed to be mutually
independent and Gaussian.
\begin{equation}
  v_n \sim \mathcal{N}(0, Q_n), \quad w_n \sim \mathcal{N}(0, R_n),
\end{equation}
where $Q_n$ is the $k \times k$ system noise covariance matrix, and $R_n$ is the $\ell
\times \ell$ observation noise covariance matrix.

\subsubsection{Dynamic Factor Model}
\label{subsubsection:dynamic_factor_model}
The dynamic factor model is a specialized class of the multivariate state-space model
designed to explain a high-dimensional observation vector using a significantly smaller
number of latent variables ($k \ll \ell$).  It can be interpreted as a time-series
extension of the static factor model.

Simply applying a standard dynamic factor model to astronomical data does not guarantee
that the estimated latent variables are uniquely associated with each physical component.
This difficulty arises from the inherent rotation indeterminacy of factor models; for
any non-singular matrix $T$, the transformations $H^* = HT$ and $x_n^* = T^{-1}x_n$
yield an identical observation equation, meaning that arbitrary linear combinations of
the states can fit the data equally well. To resolve this indeterminacy and allow the
latent variables $x_n$ to represent the assumed physical components, we introduce
strict structural constraints onto the observation matrix $H$. These constraints are
guided by prior physical insights obtained from time-averaged energy spectral fitting.
This will be explained in more detail with actual data in
\S~\ref{subsubsection:observation_model}.

\section{Data}
\label{section:observations}

\subsection{Data Acquisition and Reduction}
\label{subsection:nicer_observations_and_data_reduction} A high count rate is crucial
for the successful application of our model. We utilized the X-ray data of MAXI
J1820$+$070 obtained with NICER, which is one of the most ideal combinations to have
extreme count rates over a wide range of energy and frequency.

MAXI J$1820$+$070$ is a transient low-mass BHB discovered in March 2018 using MAXI
\citep{kawamuroMAXIGSCDetection2018}. The estimated distance is $2.96 \pm 0.33$~kpc
\citep{atriRadioParallaxBlack2020}. This proximity and a small interstellar extinction
($N_{\mathrm{H}} \sim 1.5 \times 10^{21}~\mathrm{cm}^{-2}$;
\cite{uttleyNICERObservationsMAXI2018}) resulted in a large X-ray flux over a wide
energy band. At its brightest, the source reached up to 4~Crab in the 2--6~keV energy
range \citep{shidatsuXRayOpticalNearinfrared2018}.

NICER is an X-ray timing and spectroscopy mission operational on the International Space
Station. The X-ray Timing Instrument (XTI) onboard NICER comprises 56 independent sets
of an X-ray concentrator optics and a silicon drift detector
\citep{gendreauNeutronStarInterior2016}. Beyond its primary mission objectives, NICER
serves for a broad range of astrophysical applications. Its large effective area and
wide dynamic range facilitate timing analyses at frequencies exceeding 1 kHz, a
capability previously limited in other X-ray instruments by photon shot noise or slow
response times of detectors. Additionally, NICER's moderate energy resolution (140~eV at
6~keV) and its bandpass (0.5--12~keV) enable broadband spectral analyses. All these
features make NICER particularly suited for the present study.

For data analysis, we used \texttt{HEASoft} version 6.32.1. The event files were
generated using the \texttt{NICERL2} task. We created the redistribution matrix function
(RMF) and the auxiliary response function (ARF) with the \texttt{nicerrmf} and
\texttt{nicerarf} task, respectively. The background file provided by the NICER team
was used. Because of the high brightness of MAXI J1820$+$070 during this early phase of
the outburst, the source count rate is exceptionally high (typically thousands of counts
$\mathrm{s^{-1}}$), rendering the background contribution negligible across
all analyzed energy bands and ensuring a high dynamic range for our timing analysis.
Following the advice by the NICER team
\footnote{See \url{https://heasarc.gsfc.nasa.gov/docs/nicer/data_analysis/nicer_analysis_tips.html} for details.},
events taken with the focal plane modules 14 and 34 were excluded as being noisy.

\subsection{Energy Spectral Analysis}
\label{subsection:spectral_analysis}

\begin{figure}
 \begin{center}
 \includegraphics[width=\columnwidth]{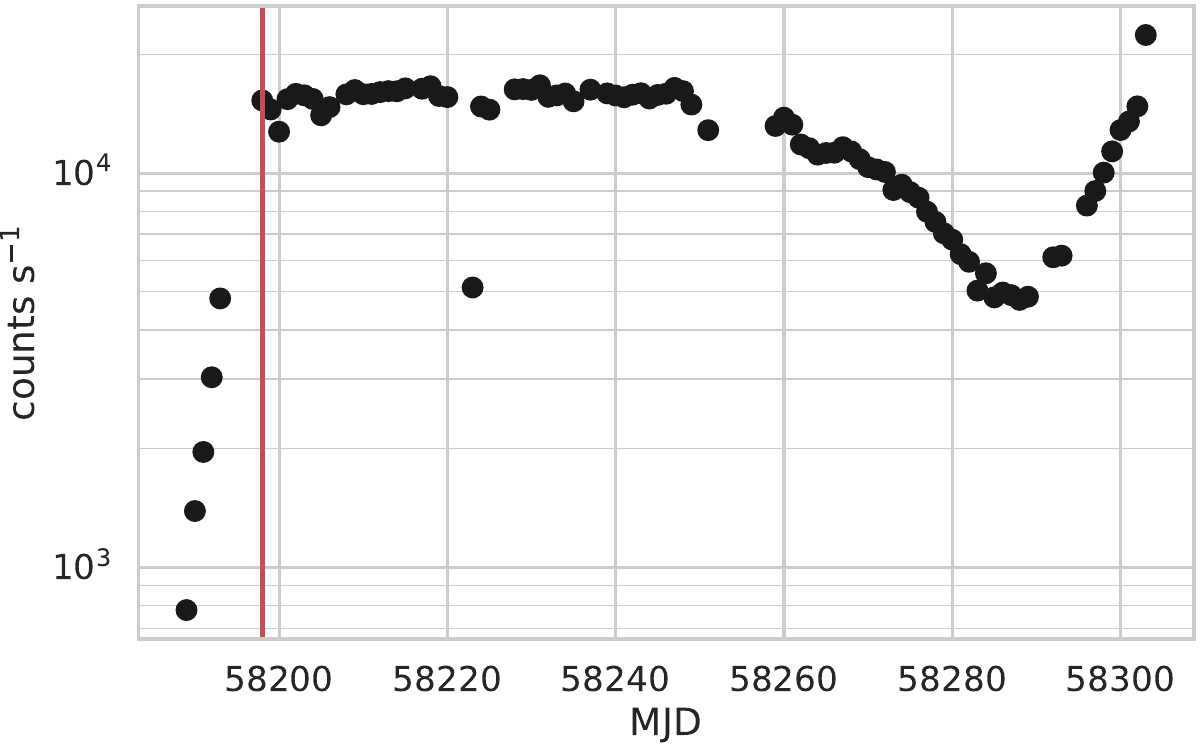}
 \end{center}
 \caption{X-ray light curve of MAXI J1820$+$070 in the entire energy band
 (0.5--10~keV). The red vertical line is the date when the O106 data set was obtained.
 }
 \label{fig:lightcurve_hardstate}
\end{figure}

Figure~\ref{fig:lightcurve_hardstate} shows the X-ray light curve of the entire X-ray
band (0.5--10~keV). In this study, we focus on data collected immediately after the
outburst on March 21, 2018 (highlighted by the red line in
figure~\ref{fig:lightcurve_hardstate}) for having one of the highest count rates in the
low hard state. This dataset (NICER observation sequence 1200120106), now referred to as
``O106'', has been extensively analyzed in numerous studies
(e.g. \cite{karaCoronaContractsBlackhole2019, buissonMaxiJ18200702019,
demarcoInnerFlowGeometry2021, zdziarskiAccretionGeometryHard2021,
omamaXrayTimeLag2023}), providing a rich basis for comparison with our results.

\begin{figure}
 \includegraphics[width=\columnwidth]{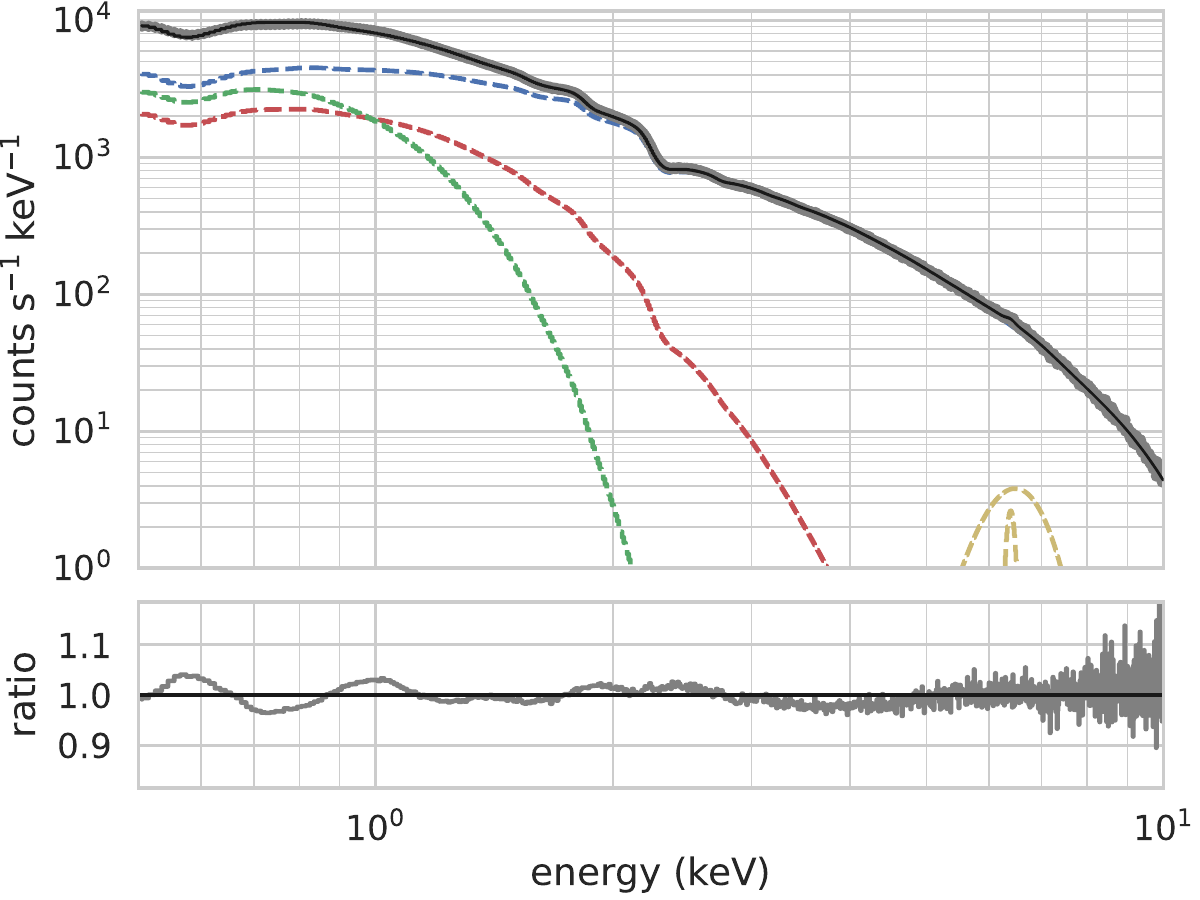}
 \caption{
 Energy spectrum of O106 with the best-fit model (upper panel) and the
 residuals (lower panel). The various components, including Comptonization
 (\texttt{powerlaw}), disk blackbody (\texttt{diskbb}), soft excess (\texttt{gauss}),
 and broad and narrow Fe lines (two \texttt{gauss} components), are represented by the
 blue, red, green, and gold curves, respectively. The residuals are shown as the
 ratio of the observed data to the model.
 }
 \label{fig:j1820_energymodel_energy_spectrum}
\end{figure}

Using the entire duration of the O106 dataset, we generated the energy spectrum
(figure~\ref{fig:j1820_energymodel_energy_spectrum}). To describe this spectrum, we
employed a phenomenological model commonly used for BHBs: \texttt{phabs * (powerlaw +
diskbb + gauss + gauss + gauss)}. The \texttt{phabs} component accounts for the
interstellar photoelectric absorption, \texttt{powerlaw} for the Comptonized emission,
\texttt{diskbb} for the multi-temperature disk blackbody emission from the accretion
disk, the first \texttt{gauss} component for the broad soft excess emission around 0.1~keV,
and the subsequent two \texttt{gauss} components are for the broad and narrow components
of the Fe K$\alpha$ emission line. The energy spectrum was analyzed within the
0.7--10~keV range, yielding a reasonably good fit. Some correlated residuals remain, but
this is good enough as the goal is not to devise an accurate spectral model,
but to determine appropriate energy bands for the band-limited light curves. The
best-fit parameters are summarized in
table~\ref{tab:j1820_energymodel_energy_spectrum_fit_results}.

\begin{table}[btp!]
 \caption{
 Fitting result of the energy spectrum fitting.
 }
  \begin{tabular}{lll}
     \hline 
     Component          & Parameter                         & Value             `                   \\
     \hline 
     \texttt{phabs}     & Column density (cm$^{-2}$)        & $0.15 \times 10^{22}$ (fixed)         \\
     \texttt{powerlaw}  & Photon index                      & $1.63 \pm0.00$                        \\
                        & Normalization (cm$^{-2}$s$^{-1}$) & $3.77 \pm 0.01$                       \\
     \texttt{diskbb}    & Temperature  (keV)                & $0.33 \pm 0.01$                       \\
                        & Normalization                     & $(2.06 \pm 0.20) \times 10^{4}$       \\
     \texttt{gaussian}  & Energy (keV)                      & $0.00 \pm 0.03$                       \\
                        & Width (keV)                       & $0.47 \pm 0.01$                       \\
                        & Normalization (cm$^{-2}$s$^{-1}$) & $18.0 \pm 2.0$                        \\
     \texttt{gaussian}  & Energy (keV)                      & $6.61 \pm 0.02$                       \\
                        & Width (keV)                       & $0.56 \pm 0.02$                       \\
                        & Normalization (cm$^{-2}$s$^{-1}$) & $0.017 \pm 0.001$                     \\
     \texttt{gaussian}  & Energy (keV)                      & $6.39 \pm 0.01$                       \\
                        & Width (keV)                       & $0.05 \pm 0.02$                       \\
                        & Normalization (cm$^{-2}$s$^{-1}$) & $0.0016 \pm 0.0003$                   \\
     \hline
  \end{tabular}
  \label{tab:j1820_energymodel_energy_spectrum_fit_results}
\end{table}

\section{Model Configuration}
\label{section:model_configuration}

\subsection{Data Definition and Band Selection}
\label{subsection:data_definition_and_band_selection}

\begin{figure}
 \includegraphics[width=\columnwidth]{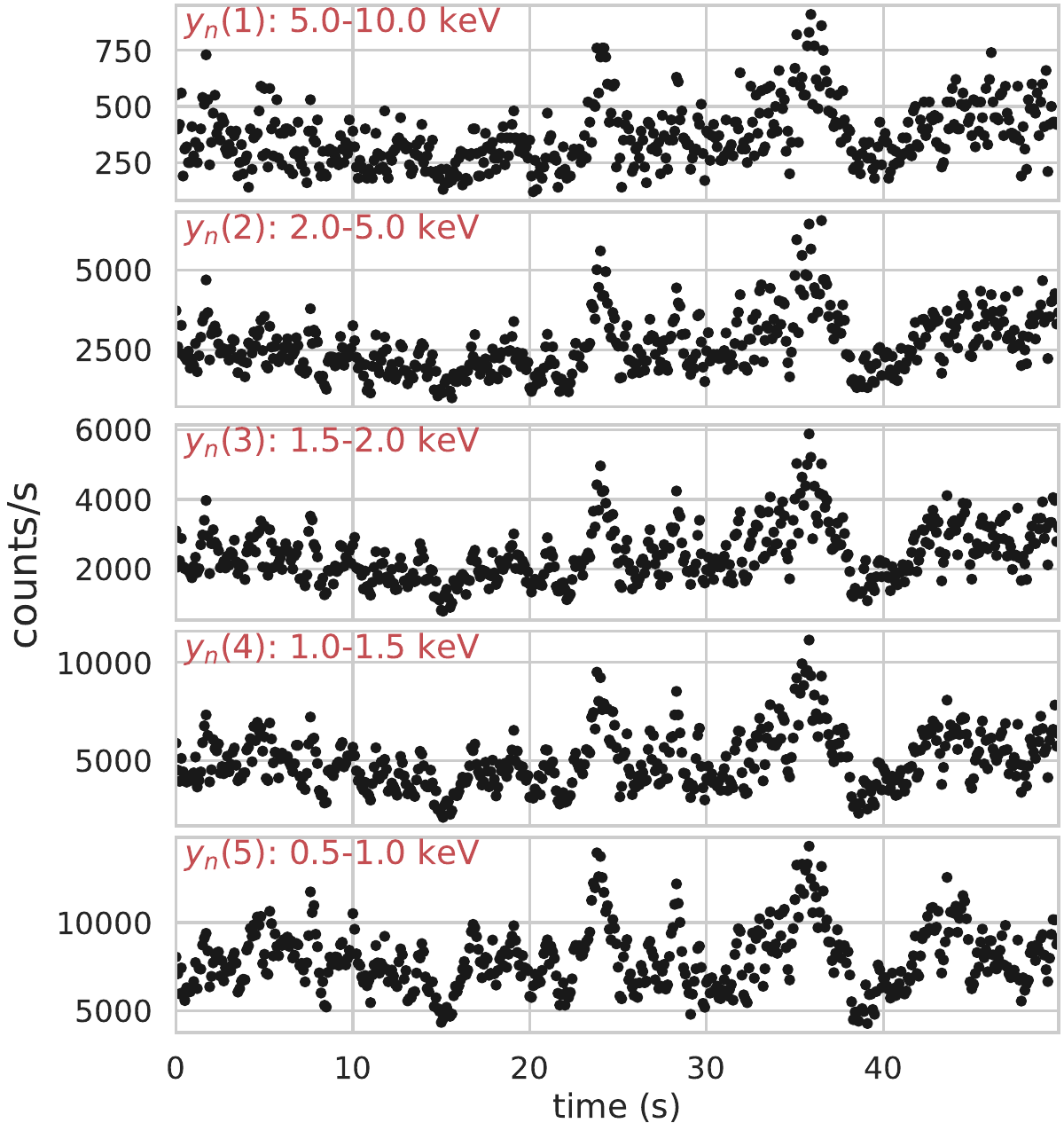}
 \caption{
  Multi-band X-ray light curves of MAXI J1820$+$070 in the 0.5--1.0, 1.0--1.5,
  1.5--2.0, 2.0--5.0, and 5.0--10.0~keV band.
 }
 \label{fig:j1820_lgssm_raw_lightcurve}
\end{figure}

We construct the energy spectrogram (Eqn.~\ref{e01}) by appropriately defining the bin
size and range of both time and energy. To satisfy the Gaussian noise assumption
required by the LGSSM, we set a time bin size of $0.1$~s so that the average count per
bin exceeds $\sim20$~counts. This bin size allows us to investigate variation below the
Nyquist frequency of $5$~Hz. This limit is significantly higher than those used in
previous studies owing to the NICER's capability to yield high count rate and the
extreme brightness of the target.

The total duration of the analyzed light curve is set to $50$~s, resulting in a total of
$N = 500$ time bins and frequency coverage down to 0.02~Hz. The duration was chosen to
be short enough to ensure that the physical system remains stationary, which is a
prerequisite for the time-invariant state-space modeling applied herein, yet long enough
to contain sufficient data that can be modeled with a reasonable computational cost.

The definition of the energy bands is directly guided by the physical contributions
identified in the time-averaged energy spectrum
(figure~\ref{fig:j1820_energymodel_energy_spectrum}). The Comptonized component
dominates across the entire energy band, whereas the other two broad components exhibit
localized contributions: the disk blackbody component contributes up to $\sim 5$~keV,
while the soft excess component is mostly confined to energies below $\sim 2$~keV.

In addition to these spectral considerations, we also consider the dimension of the
observation vector. In general, a larger number of observation variables enhances the
precision of parameter estimation. Because the photon statistics below $2$~keV are
significantly richer than those in the higher energy bands, a finer division in the
softer band is acceptable. Thus, we defined $M=5$ energy bands: 10.0--5.0, 5.0--2.0,
2.0--1.5, 1.5--1.0, and 1.0--0.5~keV.

\subsection{Specific Model Structure}
\label{subsection:specific_model_structure}

\subsubsection{Observation Model}
\label{subsubsection:observation_model}

The observation model relates the observation vector to the latent vector via a linear
combination. Based on the time-averaged spectral fitting, the number of physical components 
to be modeled is set to three ($k = 3$ as defined in
Section~\ref{subsubsection:linear_gaussian_state_space_model}) with an intention to
associate them to the three broad spectral components. 
We define the observed multi-band count rates as an $M$-dimensional 
observation vector $y_n = [y_n(1), y_n(2), y_n(3), y_n(4), y_n(5)]^T$ ($M = 5$),
where the elements $y_n(1)$ to $y_n(5)$ correspond to the energy bands of
10.0--5.0, 5.0--2.0, 2.0--1.5, 1.5--1.0, and 1.0--0.5~keV, respectively.
The observation model is expressed as:
\begin{equation}
  y_n = H_1 x_n + w_n,
  \label{eq:lgssm_j1820}
\end{equation}
where $H_1$ is the $5 \times 3$ observation matrix, and $w_n$ is the multivariate Gaussian observation noise:
\begin{eqnarray}
  H_1 &=
  \begin{bmatrix}
    h_{11} & h_{12} & h_{13} \\
    h_{21} & h_{22} & h_{23} \\
    h_{31} & h_{32} & h_{33} \\
    h_{41} & h_{42} & h_{43} \\
    h_{51} & h_{52} & h_{53}
  \end{bmatrix}.
  \\
  w_n &\sim \mathcal{N}(0, L_{R} L_{R}^T).
\end{eqnarray}
Here, $x_n = [x_n(1), x_n(2), x_n(3)]^T$ is the latent vector, whose elements represent
the Comptonization, disk blackbody, and soft excess components, respectively.
Each matrix element $h_{ij}$ denotes the structural contribution of the $j$-th latent variable to the $i$-th energy band, effectively acting as a spectral mixing matrix.
The Cholesky factor of the observation noise covariance matrix, $L_R$, is parameterized as a lower triangular matrix:
\begin{equation}
  L_{R} = 
  \begin{bmatrix}
    r_{6} & 0   & 0   & 0   & 0   \\
    r_{11} & r_{12} & 0   & 0   & 0   \\
    r_{15} & r_{14} & r_{13} & 0   & 0   \\
    r_{10} & r_{9} & r_{8} & r_{7} & 0   \\
    r_{5} & r_{4} & r_{3} & r_{2} & r_{1} \\
  \end{bmatrix}
  \label{eq:j1820_lgssm_observation_noise_covariance}
\end{equation}
where the parameter indices are arranged in a clockwise spiral for programmatic convenience.
Each matrix element $h_{ij}$ represents the structural contribution of the $j$-th latent variable to the $i$-th energy band, effectively acting as a spectral mixing matrix.

Because an unconstrained observation matrix $H_1$ is subject to arbitrary rotational
transformations, a unique solution cannot be obtained without additional structural
constraints. To break this mathematical degeneracy and allow the latent variables $x_n$ 
to represent the assumed physical components, we introduce strict structural constraints 
onto the observation matrix $H_1$. These constraints are guided by the results of the 
spectral fitting (\S~\ref{subsection:spectral_analysis}); specifically, that the soft 
excess and disk blackbody emissions make little contribution in the harder energy bands. 
Consequently, the observation matrix $H_1$ is constrained to a lower-trapezoidal structure:
\begin{equation}
  H_1 =
  \begin{bmatrix}
    h_{4} & 0   & 0   \\
    h_{6} & h_{5} & 0   \\
    h_{3} & h_{2} & h_{1} \\
    h_{7} & h_{8} & h_{9} \\
    h_{10} & h_{11}& h_{12}
  \end{bmatrix},
  \label{eq:j1820_lgssm_loading_matrix}
\end{equation}
where elements set to 0 mean that the contribution by the corresponding spectral
component is neglected in the corresponding energy band.

\subsubsection{System Model}
\label{subsubsection:system_model}

To describe the temporal evolution of the latent variables, we employ a vector
autoregressive (VAR) model \citep{lutkepohlNewIntroductionMultiple2005}.
Because a first-order VAR model ($\text{VAR}(1)$) is mathematically too restrictive to
capture local features observed in multiband light curves and slopes in the power
spectral density, we adopt a second-order VAR ($\text{VAR}(2)$) formulation to balance
flexibility and simplicity.

The $\text{VAR}(2)$ model for the core latent process $z_n$ is expressed as:
\begin{equation}
  z_n = A_1 z_{n-1} + A_2 z_{n-2} + v_n,
\end{equation}
where the coefficient matrices $A_i$ ($i = 1, \dots, p$, where $p$ denotes the VAR order and is set to $2$) are defined as:
\begin{equation}
  A_{i} =
  \begin{bmatrix}
    a_{i, 11} & a_{i, 12} & a_{i, 13} \\
    a_{i, 21} & a_{i, 22} & a_{i, 23} \\
    a_{i, 31} & a_{i, 32} & a_{i, 33} \\
  \end{bmatrix},
  \label{eq:lgssm_var_coefficients}
\end{equation}
and the system noise follows a multivariate standard Gaussian distribution:
\begin{equation}
  v_n \sim \mathcal{N}(0, Q), \quad Q = I.
\end{equation}
Here, the system noise covariance matrix $Q$ is fixed to the identity matrix $I$ to ensure model scale identifiability, absorbing the scaling factors entirely into the observation matrix $H_1$. To integrate this process into the LGSSM framework, the system model is rewritten in its companion state-space form:
\begin{equation}
  \begin{bmatrix}
  z_n   \\
  z_{n-1} \\
  \end{bmatrix}
  = 
  \begin{bmatrix}
  A_1  & A_2 \\
  I   & O  \\
  \end{bmatrix}
  \begin{bmatrix}
  z_{n-1} \\
  z_{n-2} \\
  \end{bmatrix}
  +
  \begin{bmatrix}
  v_n   \\
  O    \\
  \end{bmatrix},
\end{equation}
where the augmented state vector is defined as $x_n \equiv [z_n, z_{n-1}]^T$.

\subsection{Prior Distributions}
\label{subsection:prior_distributions}

For parameter estimation, we adopt the Bayesian inference approach, which requires
specifying prior distributions and executing a Markov Chain Monte Carlo (MCMC)
sampling. As established above, the dimensions of the latent variables, observation
variables, and the VAR model order are configured as $k = 3$, $M = 5$, and $p = 2$,
respectively.

To derive the joint posterior distribution of the model parameters, appropriate prior
distributions must be specified by considering the mathematical constraints and physical
characteristics of each parameter set. The specific choices and rationales are detailed
below for each set.

\subsubsection{VAR Matrix}
The VAR coefficient matrices are denoted as $A_i$ ($i = 1, \dots, p$) in
Equation~\eqref{eq:lgssm_var_coefficients}. Each matrix contains $k \times k$
parameters, yielding a total of $p \times k \times k = 18$ parameters to be
estimated. Physically and mathematically, these coefficients must satisfy the
stationarity condition of multivariate autoregressive processes to prevent unphysical,
explosive behavior \citep{lutkepohlNewIntroductionMultiple2005}.

Ideally, one would construct a truncated prior distribution that strictly enforces this stationarity constraint.
However, analytically evaluating the boundaries of such a stationarity region is computationally prohibitive in a high-dimensional parameter space.
In this study, we use the standard normal distribution $\mathcal{N}(0,1)$ as a weakly informative prior for each element $a_{i, jk}$.
Under non-stationary parameter regimes, the latent variable trajectories exhibit rapid exponential growth or decay, which yields an extremely low likelihood given our observed stationary time-series data.
Consequently, parameter combinations that violate the stationarity condition are naturally rejected during the MCMC sampling process.

\subsubsection{Observation Matrix}

The structurally constrained observation matrix $H_1$ in Equation~\eqref{eq:j1820_lgssm_loading_matrix} contains $Mk - k(k-1)/2 = 12$ free parameters.
Because the state variables represent physical components that contribute additively to the observed multi-band count rates, these parameters must be strictly positive ($h_i > 0$).
To enforce this positivity constraint while remaining weakly informative, we use the Half-Cauchy distribution with a scale parameter of $5$ as the prior distribution for the non-zero elements of $H_1$.

\subsubsection{Observation Noise}

For the scale of the observation noise, we similarly employ the Half-Cauchy distribution
with a scale parameter of $5$ as a weakly informative prior. To model the underlying
correlation of the multivariate observation noise, we utilize the
Lewandowski-Kurowicka-Joe (LKJ) distribution
\citep{lewandowskiGeneratingRandomCorrelation2009}.  The LKJ distribution is a standard
and computationally robust choice for prior formulations over correlation matrices in
modern Bayesian modeling.  The shape parameter $\eta$ of the LKJ distribution controls
the concentration of the probability density around the identity matrix; $\eta = 1$
represents a uniform distribution over all valid correlation matrices, while higher
values increasingly penalize strong correlations.  To incorporate a conservative prior
assumption favoring mutual independence among the energy bands while allowing the data
to drive the correlation structure, we set the LKJ shape parameter to $\eta = 3$.

\subsection{MCMC Sampling Configuration}
\label{subsection:mcmc_sampling_configuration}

To simulate the joint posterior distribution of the parameters, we execute MCMC
sampling. We implement the No-U-Turn Sampler (NUTS;
\cite{hoffmanNoUturnSamplerAdaptively2014}), an advanced variant of the Hamiltonian
Monte Carlo method for an efficient exploration of complex target distributions. The
simulation is configured with a burn-in phase of $1,000$ iterations to ensure
convergence, followed by $5,000$ production iterations for posterior inference.
Inspections of the estimated model are annexed in the appendix, including the posterior
distribution of all parameters and the model being stationary.

\section{Results}
\label{section:results}

\subsection{Reconstructed Light Curves}
\label{subsection:reconstructed_light_curves}

\begin{figure}
 \includegraphics[width=\columnwidth]{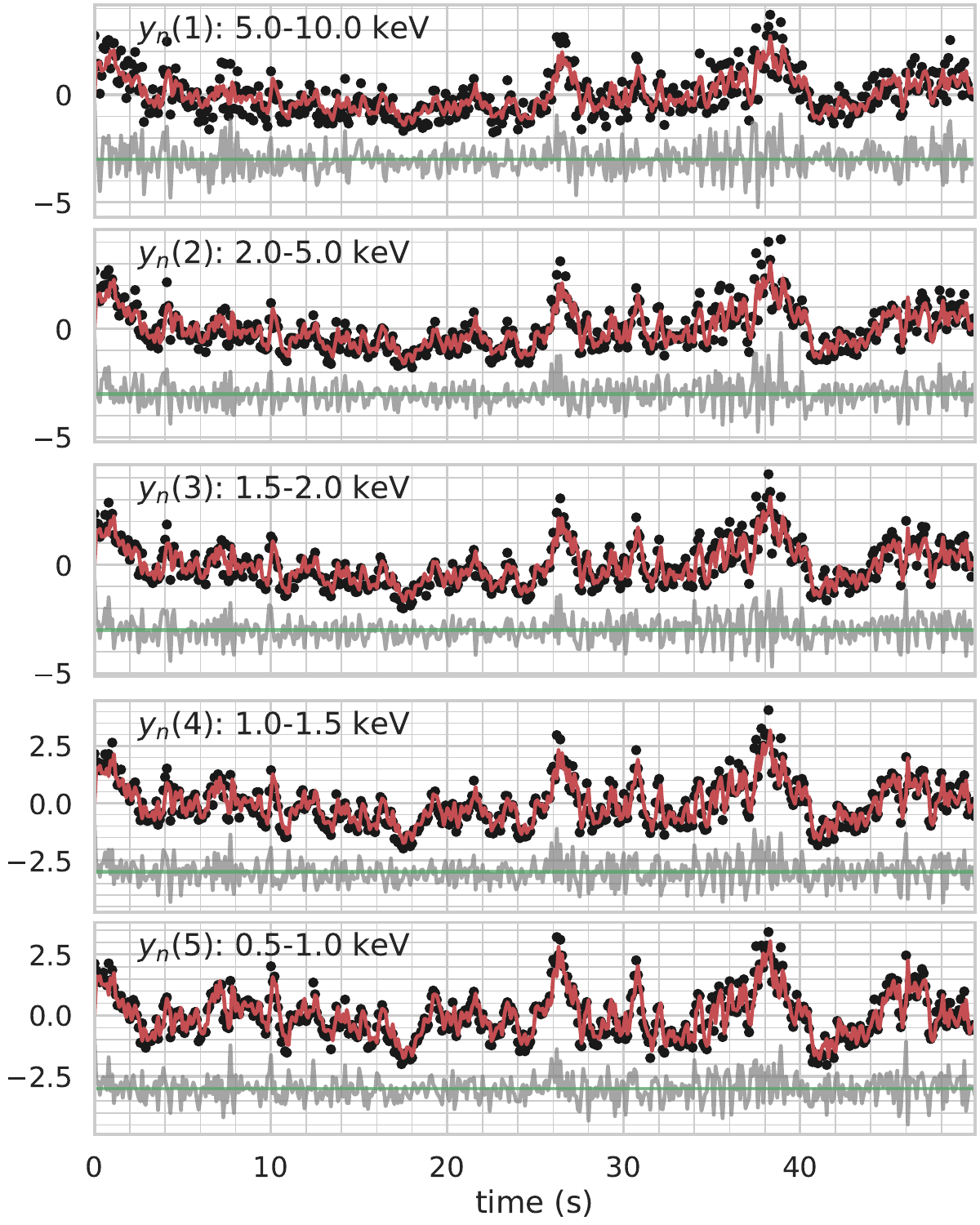}
 \caption{
  Comparison between the observed multi-band data and the model-estimated light curves for the multiband observation variables.
  In each panel, the upper portion shows the estimated light curves (red curves) overlaid on the observed data (black points), and the lower portion displays the corresponding residuals shifted downward by $-3$ for clarity.
  The shaded bands denote the 2.5\% and 97.5\% quantile credible intervals.
 }
 \label{fig:j1820_lgssm_lightcurve_observation}

 \vspace{3ex}

 \includegraphics[width=\columnwidth]{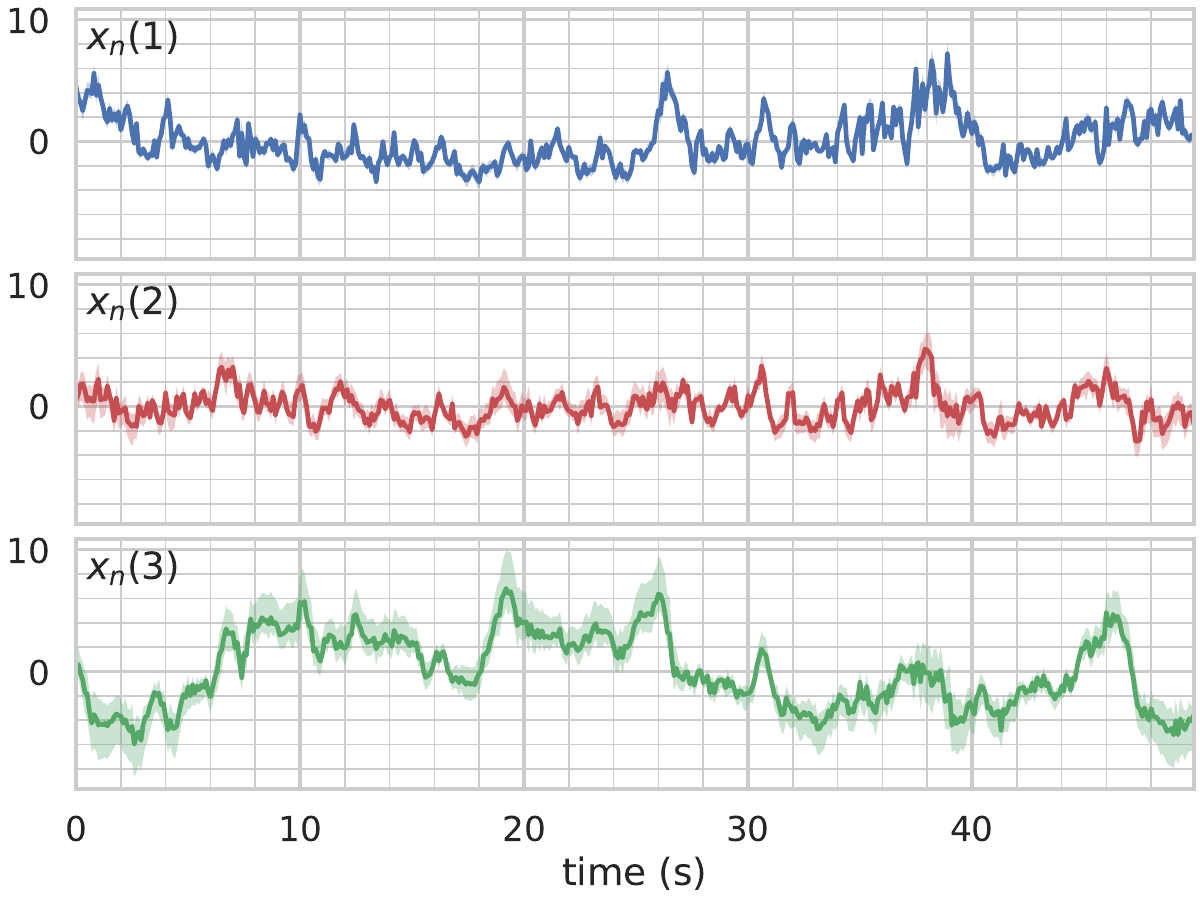}
 \caption{Estimated light curves of the latent variables representing the
 intrinsic variations of assumed physical components. The solid curves show the
 median values and their shaded areas indicate the 2.5\% and 97.5\% quantile credible
 intervals.
 }
 \label{fig:j1820_lgssm_lightcurve_smoothed_state}
\end{figure}

For each MCMC sample drawn from the joint posterior distribution, the K\'{a}lman
smoothing algorithm is applied to reconstruct the continuous time-series profiles of
both the observation (figure~\ref{fig:j1820_lgssm_lightcurve_observation}) and latent
(figure~\ref{fig:j1820_lgssm_lightcurve_smoothed_state}) variables. 

As in the upper panels of figure~\ref{fig:j1820_lgssm_lightcurve_observation}, the
smoothed estimates for the observation variable reproduce both the overall variability
and rapid stochastic fluctuations across all five energy bands reasonably well.  The
corresponding residuals in the lower panels display no significant coherent structures,
validating the structural adequacy of the model. The smoothing algorithm extracts the
temporal evolution of the latent variables, thereby yielding individual, decomposed
light curves for each physical component
(figure~\ref{fig:j1820_lgssm_lightcurve_smoothed_state}). This is the advantage of this
approach, offering a more robust representation against spectral mixture compared to
traditional timing techniques or time-averaged spectral fitting alone.
Figure~\ref{fig:j1820_lgssm_lightcurve_smoothed_state} demonstrates the utility of the
state-space model in evaluating the time-domain profiles of the assumed physical components.

In figure~\ref{fig:j1820_lgssm_lightcurve_smoothed_state}, the smoothed light curves
reveal distinct variability characteristics unique to each latent variable.  The latent
variables for Comptonization ($x_n(1)$) and the disk blackbody ($x_n(2)$) are heavily
dominated by rapid, high-frequency fluctuations throughout the duration.  In contrast,
the latent variable for the soft excess ($x_n(3)$) prominently exhibits lower-frequency
modulations of a larger amplitude. A notable morphological feature is captured at
approximately $46$~s, where a distinct low-frequency dip occurs exclusively in $x_n(3)$.
This behavior is consistent with the multi-band raw data
(figure~\ref{fig:j1820_lgssm_lightcurve_observation}), where this localized
low-frequency modulation becomes increasingly pronounced in the lower energy channels
where the soft excess is assumed to dominate.

\subsection{Frequency-Domain Dynamics}
\label{subsection:frequency_domain_dynamics}

\begin{figure}
 \includegraphics[width=\columnwidth]{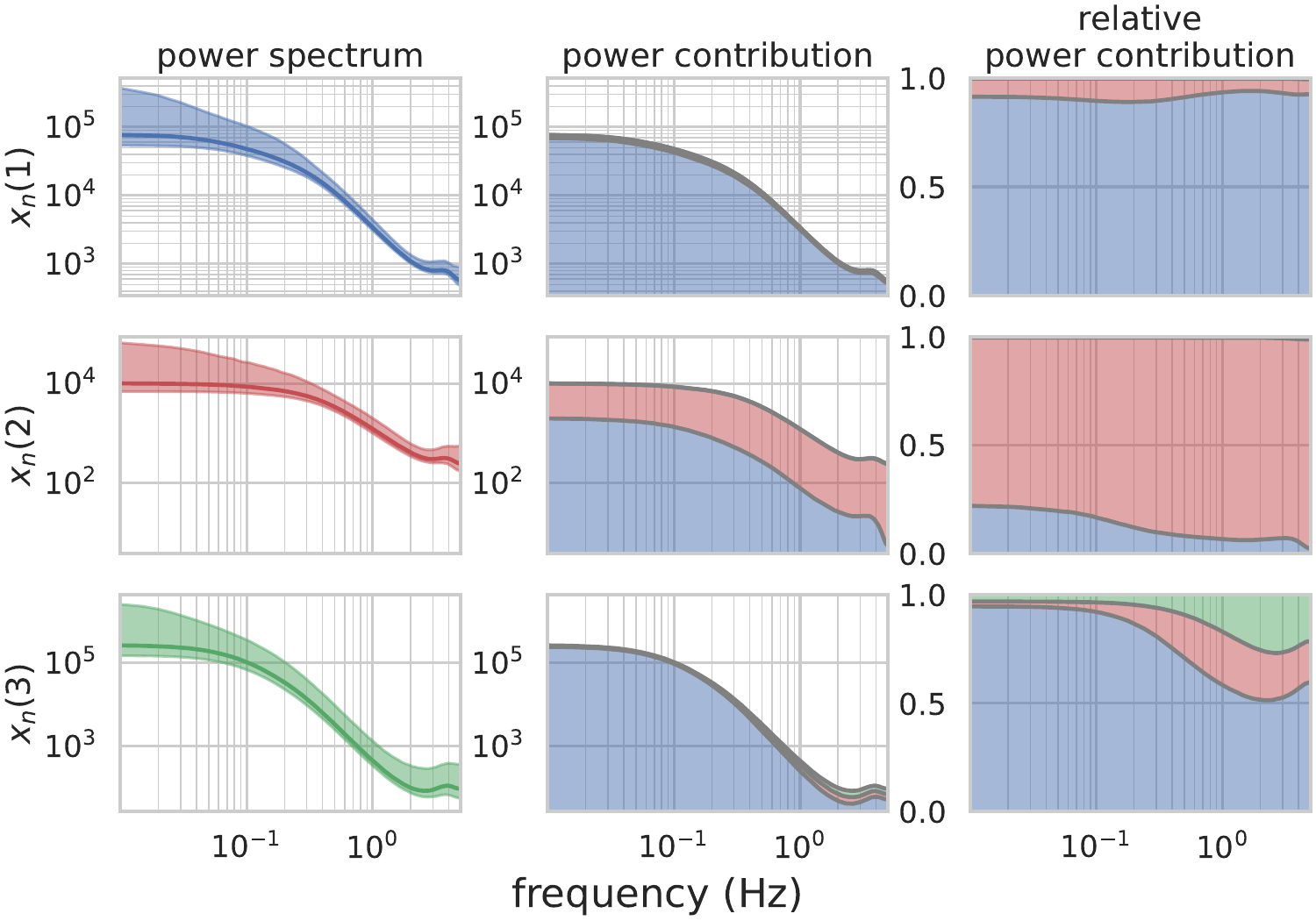}
 \caption{
  Absolute and relative power contributions estimated from the VAR (vector autoregressive) matrix in the frequency domain.
  The rows from top to bottom correspond to the target latent variables: Comptonization ($x_n(1)$), disk blackbody ($x_n(2)$), and soft excess ($x_n(3)$), respectively.
  The blue, red, and green shaded regions represent the dynamic power contributions originating from $x_n(1)$, $x_n(2)$, and $x_n(3)$, with the bands indicating the 95\% quantile credible intervals.
 }
 \label{fig:j1820_lgssm_power_contribution}
\end{figure}

The estimated VAR coefficient matrices encapsulate the intrinsic dynamic properties of
the underlying physical system. By transforming these multivariate coefficients into the
frequency domain, we can evaluate the frequency-domain system functions---specifically,
the power spectral densities (PSDs) and cross-coherences of the latent variables---for the
system's underlying behavior. These functions reflect the analytical properties of the
estimated VAR model, which are distinct from the empirical statistics of the raw
multi-band light curves. We show that the utility of this approach extends to
evaluating the coupled dynamics and correlations embedded within the latent variables
robustly against spectral mixture.

To perform a physically meaningful comparison of the PSDs among the
different components, the latent variables must be projected onto a common flux scale.
This is achieved by rescaling the variance of the estimated states into the scale of the
softest energy band ($0.5$--$1.0$~keV), following the procedure detailed later in
\S~\ref{subsection:reconstruction_of_component_energy_spectra}.
This specific band is selected because all three assumed physical components exhibit
significant contributions to the total photon counts, allowing their relative
variability amplitudes to be directly compared.

The leftmost panels in figure~\ref{fig:j1820_lgssm_power_contribution}
present the PSDs of the latent components.
The overall PSD profiles of all three components exhibit a standard stochastic
continuum: a flat plateau at low frequencies, a power-law slope above a specific break
frequency, and an excess power structure at high frequencies.  However, the
characteristic break frequencies differ noticeably among the components; the
Comptonization, disk blackbody, and soft excess components exhibit breaks at
approximately $0.05$, $0.1$, and $0.05$~Hz, respectively. Consequently, the soft excess
component displays a larger relative power contribution in the low-frequency regime,
providing a quantitative explanation for why low-frequency modulations visually dominate
the softer-band light curves (figure~\ref{fig:j1820_lgssm_raw_lightcurve}). In the
high-frequency regime, all three components exhibit a synchronized excess power signal
centered around $4.0$~Hz. Because it is physically improbable for three spatially
distinct emission regions to spontaneously generate independent periodic oscillations at
an identical frequency, this coherence suggests that a temporal signal propagates
dynamically from one component to the others.

To quantify these correlations, we evaluate the power contribution functions computed
from the estimated VAR coefficient matrices
(figure~\ref{fig:j1820_lgssm_power_contribution}).  These functions decompose the total
PSD of each variable into the relative shares originating from the
individual noise terms across frequencies (for details, see
\cite{kitagawaIntroductionTimeSeries2022}).  For the relative power profiles of the
Comptonization ($x_n(1)$) and disk blackbody ($x_n(2)$) components, the
self-contribution is highly dominant across all frequencies. Conversely, the soft excess
component ($x_n(3)$) presents a frequency-dependent contributions: its low-frequency
variance is predominantly composed of the contribution of the Comptonization component
($x_n(1)$). This external contribution decreases at higher frequencies, where the
self-contribution of $x_n(3)$ and the contribution from $x_n(2)$ gradually increase. For
the disk component, the relative contribution from the Comptonization component
similarly exhibits a monotonic decrease with frequency. The $4.0$-Hz feature does not
produce a sharp peak in the relative power contribution, indicating that it is embedded
within the broadband continuum propagation.

\subsection{Time-Domain Dynamics}
\label{subsection:time_domain_dynamics}

\begin{figure}
 \includegraphics[width=\columnwidth]{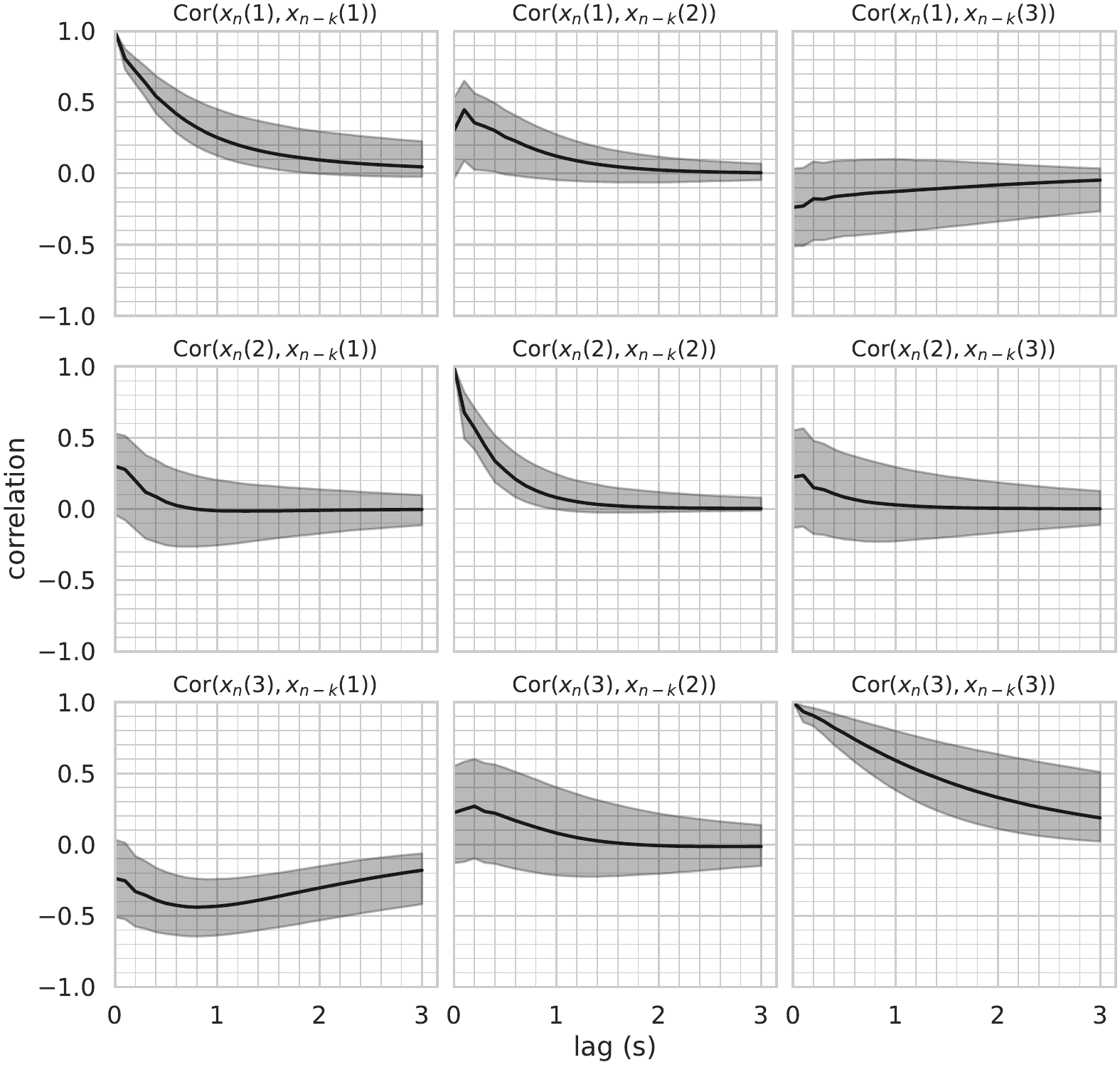}
 \caption{
  Cross-correlation functions derived from the estimated VAR model among the latent variables.
  The indices $i$ and $j$ denote the row and column coordinates of the panel matrix following Equation~\eqref{eq:cross_correlation_function}, with the diagonal panels representing the autocorrelation profiles.
  The variables $x_n(1)$, $x_n(2)$, and $x_n(3)$ represent the latent variables for the assumed Comptonization, disk blackbody, and soft excess components, respectively.
  The shaded regions denote the 95\% posterior credible intervals.
 }
 \label{fig:j1820_lgssm_crosscorrelation}
\end{figure}

To investigate the correlations in the time domain, the cross-correlation functions
among the latent variables are computed directly from the estimated VAR coefficient
matrices (figure~\ref{fig:j1820_lgssm_crosscorrelation}). These functions represent the
correlation structure of the latent variables themselves separated from empirical sample
fluctuations and spectral mixing. The
cross-correlation function between two latent variables is defined as:
\begin{equation}
 C_k(i, j) = \mathrm{Cor}(x_n(i), x_{n-k}(j)),
 \label{eq:cross_correlation_function}
\end{equation}
where a positive lag $k$ indicates that the signal in state $i$ lags behind the signal
in state $j$. Figure~\ref{fig:j1820_lgssm_crosscorrelation} shows the result for all combinations.

We first focus on the diagonal panels, which represents the auto-correlation of the
latent variables. The distinctive differences in the break frequencies observed in the
PSD profiles (Section~\ref{subsection:frequency_domain_dynamics}) are
clearly reflected in the temporal widths of their respective autocorrelation
functions. Specifically, the soft excess component, $C_k(3, 3)$, exhibits the broadest
temporal profile reflecting its low-frequency dominance. Conversely, the
Comptonization ($C_k(1, 1)$) and disk blackbody ($C_k(2, 2)$) components display
significantly narrower widths, consistent with their high-frequency stochastic
characteristics.

We next focus on the off-diagonal panels. A delayed signal is resolved
in the cross-correlation between the Comptonization and disk blackbody components,
$C_k(1, 2)$, exhibiting a well-defined correlation peak at a lag of $0.1$~s. Similarly,
the cross-correlation between the soft excess and the disk blackbody, $C_k(3, 2)$,
reveals a time-lagged correlation peaking at a lag of $0.2$~s. While both propagation
pathways originate from the intrinsic variance of the disk blackbody ($x_n(2)$), the
response of the soft excess is delayed by an additional $0.1$~s relative to the
Comptonization component. This systematic temporal progression suggests a
signal propagation sequence structured as $x_n(2) \rightarrow x_n(1)
\rightarrow x_n(3)$.

In contrast, in the cross-correlation between the soft excess and the Comptonization component, $C_k(3, 1)$, the direct forward propagation from $x_n(1)$ to $x_n(3)$ is obscured by an anti-correlation.
Specifically, the soft excess component exhibits an opposite-sign fluctuation approximately $0.5$~s after a given variation occurs in the Comptonization component.
This delayed anti-correlation accounts for the dominance of the Comptonization
component ($x_n(1)$) in the low-frequency power contribution of
the soft excess ($x_n(3)$), providing a consistent time-domain explanation
for the coupled behavior observed in the frequency-domain analysis.

\subsection{Reconstruction of Component Energy Spectra}
\label{subsection:reconstruction_of_component_energy_spectra}

Conversely, the successful estimation of the variability of the latent variables
indicates that a spectral decomposition can be achieved via time-series analysis, which
we demonstrate here. Because the system model assumes a $\text{VAR}(2)$ process with a
zero mean, the latent variables $x_n$ have a theoretical time-averaged expectation of
zero.  Consequently, the time-averaged energy spectrum for each assumed physical
component can be reconstructed by projecting the estimated factor loadings back onto the
original scale of the multi-band light curves. To reverse the standardization applied during data
pre-processing, the reconstructed count rates for the $j$-th latent variable in the
$i$-th energy band, denoted as $c_{ij}$, are computed via the following vector
transformation:
\begin{equation}
  \begin{bmatrix}
    c_{1j} \\
    c_{2j} \\
    c_{3j} \\
    c_{4j} \\
    c_{5j} \\
  \end{bmatrix}
  =
  \begin{bmatrix}
    \sigma_{1} & 0 & 0 & 0 & 0 \\
    0 & \sigma_{2} & 0 & 0 & 0 \\
    0 & 0 & \sigma_{3} & 0 & 0 \\
    0 & 0 & 0 & \sigma_{4} & 0 \\
    0 & 0 & 0 & 0 & \sigma_{5} \\
  \end{bmatrix}
  \begin{bmatrix}
    h_{1j} \\
    h_{2j} \\
    h_{3j} \\
    h_{4j} \\
    h_{5j} \\
  \end{bmatrix}
  +
  \begin{bmatrix}
    \mu_{1} \\
    \mu_{2} \\ 
    \mu_{3} \\ 
    \mu_{4} \\ 
    \mu_{5} \\
  \end{bmatrix},
  \label{eq:j1820_lgssm_rescaling}
\end{equation}
where $j \in \{1, 2, 3\}$ is used for the latent variables representing the assumed physical
components (Comptonization, disk blackbody, and soft excess, respectively), and $h_{ij}$
denotes the element of the constrained observation matrix $H_1$ corresponding to the
$i$-th energy band and the $j$-th state. The parameters $\mu_{i}$ and $\sigma_{i}$
represent the empirical time-averaged mean and standard deviation of the observed
multi-band light curve in the $i$-th band, respectively.

\begin{figure}
 \includegraphics[width=\columnwidth]{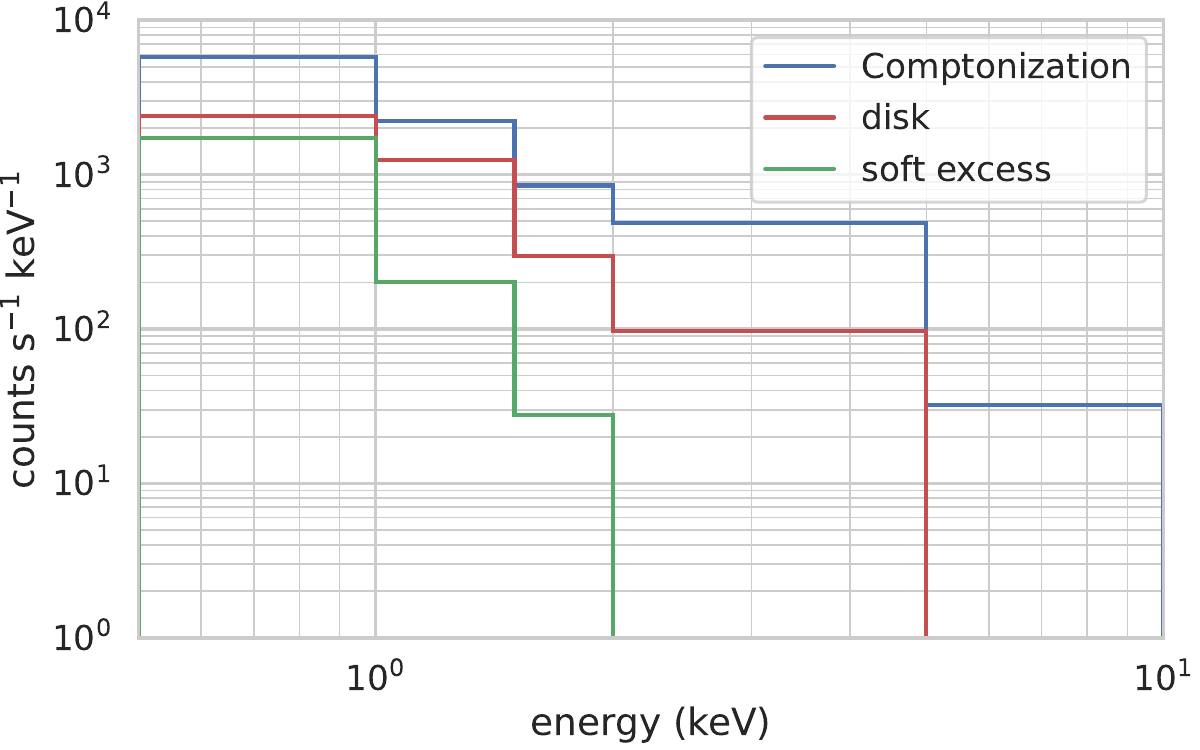}
 \caption{
  Energy spectra of the assumed physical components reproduced from the estimated observation matrix.
  The spectral profiles of the latent variables associated to the Comptonization, disk
  blackbody, and soft excess components are reconstructed based on their factor loadings.
 }
 \label{fig:j1820_lgssm_reproduced_energy_spectrum}
\end{figure}

The resultant reconstructed time-averaged energy spectrum for each assumed physical
component (figure~\ref{fig:j1820_lgssm_reproduced_energy_spectrum}) closely resembles
the energy spectrum model independently obtained through standard time-averaged energy
spectral fitting (figure~\ref{fig:j1820_energymodel_energy_spectrum}).  It is important
to emphasize that this spectral decomposition is achieved solely through multivariate
time-series modeling. While the configuration utilizes the prior spectral insights to
define the structured zero-elements of the observation matrix $H_1$
(Equation~\eqref{eq:j1820_lgssm_loading_matrix}), it requires no prior input regarding
the specific analytical shape of the energy spectrum.  This agreement supports the
physical relevance of the estimated latent variables, demonstrating that the LGSSM
successfully provides a robust representation of the variability derived under these
physical assumptions.

\section{Discussion}
\label{section:discussion}

\subsection{Physical Interpretation of Variation Timescales}
\label{subsection:timescales}

Traditional multi-band light-curve analyses have often led to contradictory conclusions 
regarding the geometry of the accretion disk around BHBs. In case of the low/hard 
state of MAXI J1820$+$070, some studies focusing on X-ray reverberation 
lags have argued that the accretion disk extends inward close to the innermost stable 
circular orbit \citep{karaCoronaContractsBlackhole2019, buissonMaxiJ18200702019}.
Other analyses based on optical/X-ray time lags and time-averaged spectra
have suggested a truncated disk geometry
\citep{demarcoInnerFlowGeometry2021, zdziarskiAccretionGeometryHard2021, axelssonAccretionGeometryBlack2021, omamaXrayTimeLag2023}.
This discrepancy arises not only from the spectral mixture of multiple emission
components within selected energy bands but also from the empirical,
waveband-dependent nature of conventional timing metrics.
By applying the proposed LGSSM framework, we evaluated the 
variability of the latent variables representing the assumed physical components, 
enabling an investigation of their relationships with enhanced robustness against 
instrumental and energy band mixing effects.

We first evaluate the characteristic timescales of the extracted physical components
based on their PSD. As seen in \S~\ref{subsection:frequency_domain_dynamics}, the
Comptonization ($x_n{(1)}$), disk blackbody ($x_n{(2)}$), and soft excess ($x_n{(3)}$)
components exhibit distinct break frequencies at approximately 0.05~Hz ($\sim 20$~s),
0.1~Hz ($\sim 10$~s), and 0.05~Hz ($\sim 20$~s), respectively. These break frequencies
correspond to the intrinsic characteristic timescales of the underlying physical
processes governing each decoupled emission region.

To associate these timescales with underlying physical processes, we compare them with 
the fundamental timescales of the standard accretion disk model
\citep{shakuraBlackHolesBinary1973}. 
Assuming a black hole mass of $M = 10 M_\odot$, the Keplerian orbital time $t_{\rm K}$ 
at a radius $r$ (in units of the gravitational radius $r_{\rm g}$) is approximated as
\begin{equation}
t_{\rm K} \approx 0.3 \left( \frac{M}{10M_\odot} \right) \left( \frac{r}{50r_{\rm g}} \right)^{3/2} \, {\rm (s)},
\end{equation}
while the viscous timescale $t_{\rm visc}$ is given by
\begin{equation}
t_{\rm visc} \sim \alpha^{-1} \left( \frac{h}{r} \right)^{-2} \Omega_{\rm K}^{-1} = \frac{t_{\rm K}}{2\pi \alpha} \left( \frac{h}{r} \right)^{-2},
\end{equation}
where $\alpha$ is the viscosity parameter, $h$ is the disk thickness,
and $\Omega_{\rm K} \equiv 2\pi / t_{\rm K}$ is the Keplerian angular frequency.

Under the truncated disk scenario commonly applied to the low/hard state
\citep{doneModellingBehaviourAccretion2007}, the Comptonized emission is assumed to
originate from a hot inner accretion flow extending inward from the truncated disk
edge. If we associate the break timescale of $x_n{(1)}$ ($\sim 20$~s) with the viscous
timescale at the innermost radius of the thin disk, and adopt typical parameters of
$\alpha = 0.2$ and $h/r = 0.2$, we obtain an estimated truncation radius of $r \sim
100r_{\rm g}$. Concurrently, if we associate the break timescale of the disk component
$x_n{(2)}$ ($\sim 10$~s) with the Keplerian orbital or thermal timescale at the same
boundary, we independently obtain a consistent scale of $r \sim 100r_{\rm g}$. The value
also agrees with the estimates derive by other means in the literature
\citep{demarcoInnerFlowGeometry2021, zdziarskiAccretionGeometryHard2021} All together,
this supports the truncated accretion disk scenario for MAXI J1820$+$070 in the low/hard
state.

\subsection{Implications for the Accretion Disk Geometry}
\label{subsection:geometry}

The cross-correlation functions among the latent variables can reveal relationships
among them more robustly against spectral mixing compared to the conventional
cross-correlation among multi-band light curves, and they yield consistent results. We
examine the time lags between the three pairs below.

Between the disk blackbody
($x_n{(2)}$) and Comptonization ($x_n{(1)}$) components, a distinct time lag of $\sim
0.1$~s is observed between the two latent variables. This lag is known as ``hard lag'',
which is frequently detected in conventional timing analyses across different
energy bands \citep{miyamotoDelayedHardXrays1988,nowakRossiRayTiming1999}, where the
harder energy bands are dominated by the Comptonized component. The time lag is too
large for a pure light-crossing time (equivalent to $\sim 10^4 r_{\rm g}$) and is more
naturally explained by the viscous propagation model of accretion rate fluctuations within the hot flow
\citep{lyubarskiiFlickerNoiseAccretion1997,kotovXrayTimelagsBlack2001,
arevaloInvestigatingFluctuatingaccretionModel2006}.  In this context, the 0.1-s delay
corresponds to the propagation time through the innermost region of the accretion flow,
where local perturbations modulate the disk blackbody emission before propagating inward
to trigger the Comptonized emission.

Between the Comptonization ($x_n{(1)}$) and the soft excess ($x_n{(3)}$), the
cross-correlation exhibits an anti-correlation peaking at a lag of $\sim 0.5$~s.  While
some authors explore sub-millisecond scales to capture reverberation signals between
the pairs \citep{uttley2014,karaCoronaContractsBlackhole2019}, it is not accessible
with the LGSSM configuration for the current data set, in which the 0.1-s time-binning
is required to ensure the Gaussianity for the observational noise covariance.

Between the the disk blackbody ($x_n{(2)}$) and the soft excess ($x_n{(3)}$), a time lag
of $\sim 0.2$~s is detected. Combined with the power contribution analysis, which shows
that the low-frequency variance of $x_n{(3)}$ is driven predominantly by the
fluctuations originating from $x_n{(1)}$, we argue for a systematic progression of
signals among the latent variables: $x_n{(2)} \rightarrow x_n{(1)} \rightarrow x_n{(3)}$. The
observation that high-frequency fluctuations in $x_n{(1)}$ do not significantly transfer
to $x_n{(3)}$ indicates that the disk response effectively acts as a low-pass filter.
This behavior is consistent with the viscous diffusive and thermal processing nature of
a reprocessing accretion disk illuminating the outer regions
\citep{lyubarskiiFlickerNoiseAccretion1997, wilkinsonAccretionDiscVariability2009}.

\medskip

\begin{figure}
 \includegraphics[width=\columnwidth]{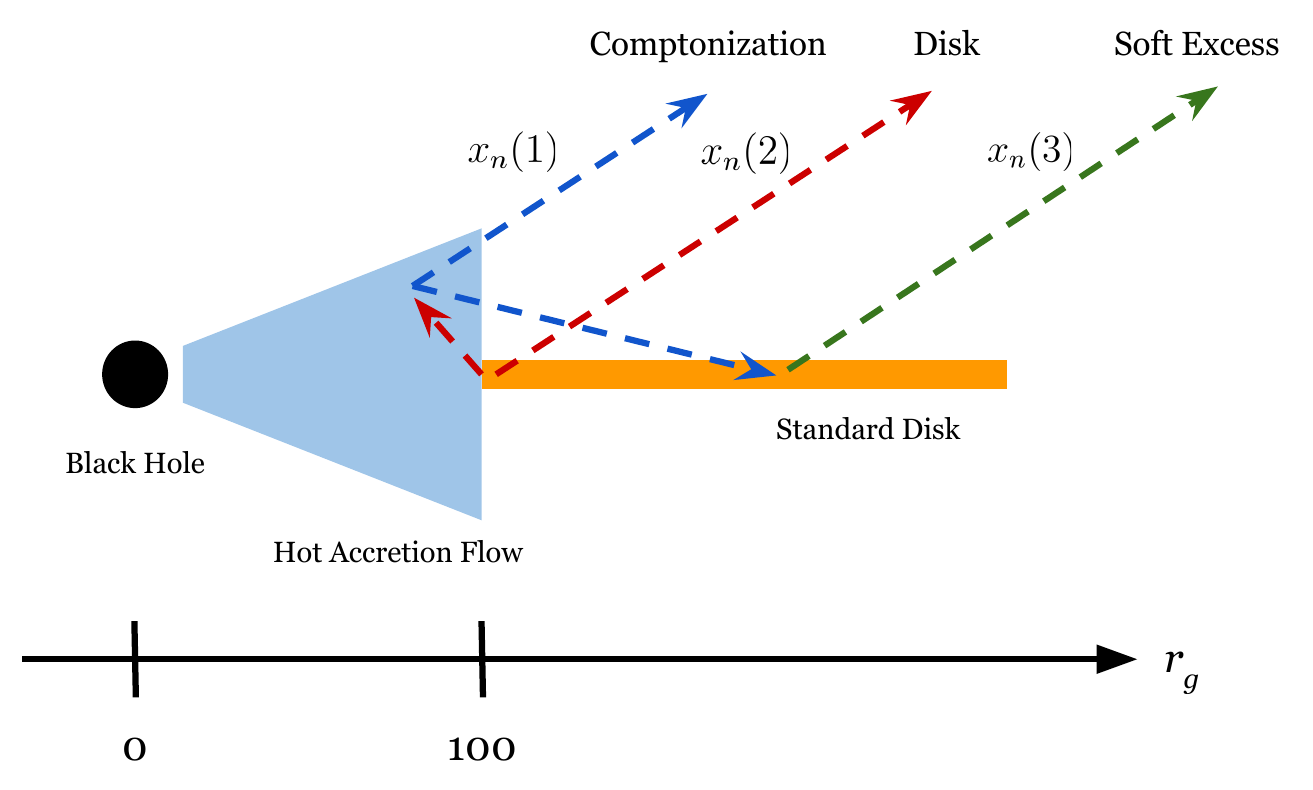}
 \caption{
 Sketch of the truncated accretion disk geometry for MAXI J1820$+$070 in the low/hard state, illustrating the physical configuration of the emission components.
 }
 \label{fig:schematics_truncate_disk}
\end{figure}

Synthesizing the variation timescales of the latent variables and the time lags between them,
we reach a geometry of MAXI J1820$+$070 depicted in
figure~\ref{fig:schematics_truncate_disk}. Our findings are compatible with the
truncated accretion disk model, which is advocated also by some authors
\citep{demarcoInnerFlowGeometry2021, zdziarskiAccretionGeometryHard2021, axelssonAccretionGeometryBlack2021, omamaXrayTimeLag2023}.
Local fluctuations in the accreting matter near the inner truncated radius ($r \sim
100r_{\rm g}$) modulate the disk blackbody emission represented by $x_n{(2)}$.  These
perturbations propagate inward into the hot inner flow or corona, producing the
Comptonized hard X-rays ($x_n{(1)}$) with a characteristic delay of $\sim 0.1$~s. These
hard X-rays subsequently illuminate the outer thin disk, where they are reprocessed and
emerge as the soft excess component ($x_n{(3)}$) with an additional thermal delay. This
demonstrates that the wealth of information obtained with the state-space model can
converge into a self-consistent picture of an astrophysical phenomenon.

\section{Summary and Outlook}
\label{section:summary_and_outlook}

In this paper, we proposed a novel joint spectral and timing modeling framework to
address the challenge of spectral mixture inherent in traditional multi-band light-curve
analyses.  By extending the dynamic factor model within an LGSSM and integrating the
structural constraints derived from time-averaged energy spectroscopy, we resolved the
inherent rotation indeterminacy of factor models.  This framework allows for a formal
decomposition of observed multi-band light curves into latent variables corresponding to
the assumed physical components, enabling the evaluation of their intrinsic variation
commererobustly against spectral mixture effects.

To demonstrate the validity and utility of this approach, we applied it to the high
count rate, multi-band X-ray light curves of the black hole binary MAXI J1820$+$070
obtained with NICER during its low/hard state.  The model successfully evaluated the
continuous, time-domain variations of three latent variables representing three
broadband spectral components: Comptonization, disk blackbody, and soft excess.  The
fact that the time-averaged energy spectra reconstructed from the factor loadings
(figure~\ref{fig:j1820_lgssm_reproduced_energy_spectrum}) closely match the results of
independent spectral fitting (figure~\ref{fig:j1820_energymodel_energy_spectrum})
supports the physical relevance of these estimated latent variables.  This structural
agreement suggests that the model effectively bridges the mathematical latent states and
the assumed physical components without statistical over-fitting.

By analyzing the system functions in both the frequency and time domains, we obtained
some clear insights into the underlying processes that were subject to some ambiguity
due to spectral mixing:
\begin{itemize}
  \item The characteristic break frequencies for the Comptonization, disk blackbody, and
	soft-excess components were evaluated to be approximately $0.05$~Hz, $0.1$~Hz,
	and $0.05$~Hz, respectively, reflecting the intrinsic timescales of each
	physical component.
  \item A clear signal propagation sequence was resolved along the path of disk
	$\rightarrow$ Comptonization $\rightarrow$ soft excess, featuring discrete time
	lags of $0.1$~s and $0.2$~s relative to the disk component.
\end{itemize}
These results supports the idea of the truncated accretion disk scenario for the MAXI
J1820$+$070 in the low/hard state, as was argued by some previous studies.

\medskip

The presented framework is generally applicable to other multi-band, high-cadence
astronomical time series data where the observed emission is a linear combination of
distinct, overlapping spectral components. Potential future applications include the
study of rapid variability in neutron star binaries, cataclysmic variables, and active
galactic nuclei, where the physical decoupling of thermal disk emission and non-thermal
corona components remains a key challenge in understanding accretion disk physics. We
can also expand the presented framework to incorporate non-Gaussian or non-linear state
equations, so that we can apply it for non-stationary astrophysical phenomena, such as
state transitions or flares.

In conclusion, this study demonstrates that the integration of physical spectral
constraints into a dynamic factor model provides a robust methodology to evaluate
overlapping variations from multi-band light curves. This framework offers a promising
approach to moving beyond purely empirical descriptions of observed data toward a
clearer physical understanding of intrinsic variability of physical components via
latent variables.

\section*{Funding}
This work was supported by JSPS KAKENHI Grant Number 22J13440.

\appendix

\section{Inspection of the Estimated Model}\label{sA1}

\begin{figure*}[htbp]
 \begin{center}
 \begin{minipage}{0.75\hsize}
  \includegraphics[width=\columnwidth]{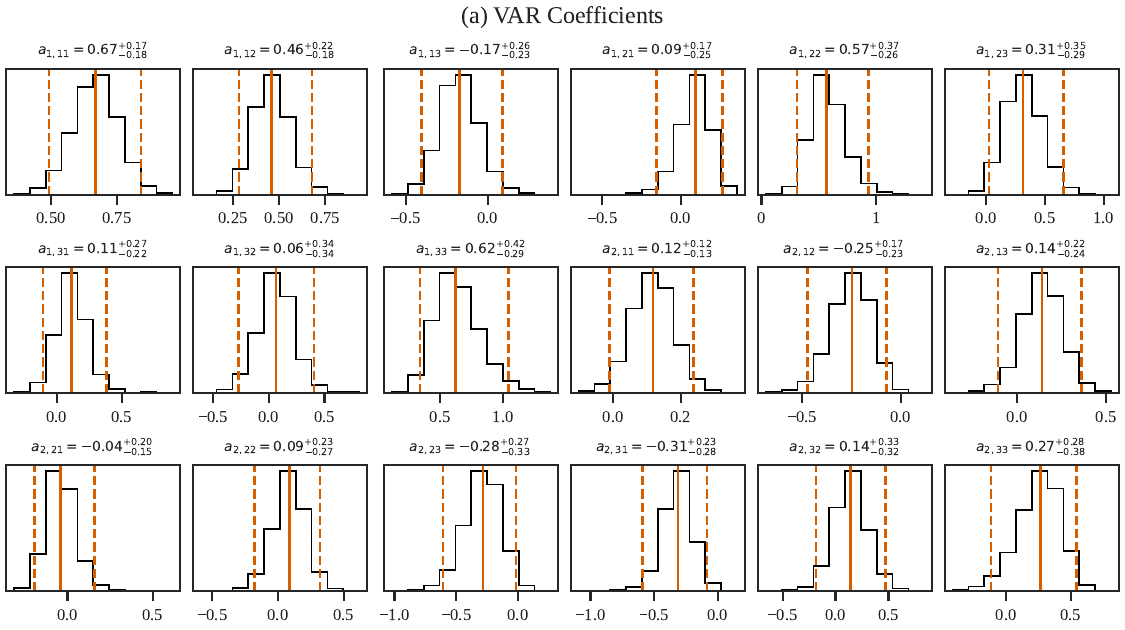}
 \end{minipage} \\
 \vspace{3ex}
 \begin{minipage}{0.75\hsize}
  \centering
  \includegraphics[width=\columnwidth]{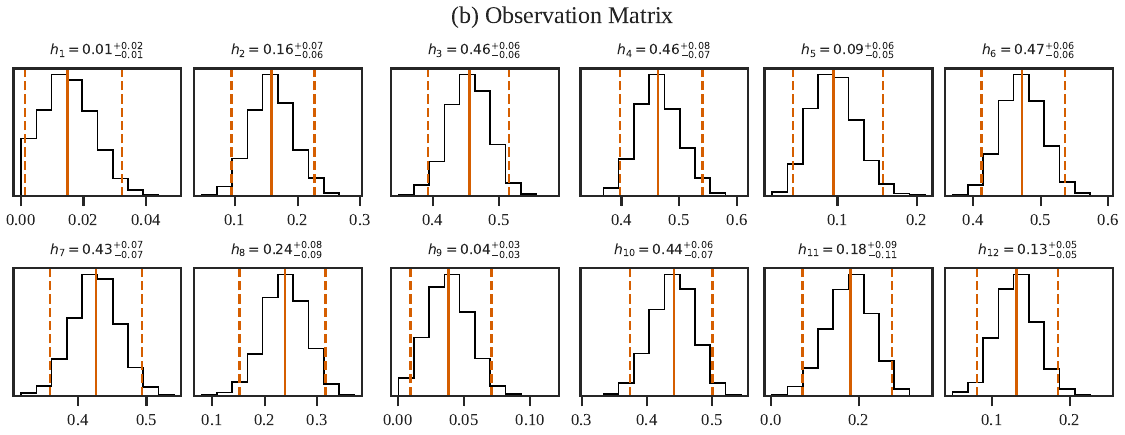}
 \end{minipage} \\
 \vspace{3ex}
 \begin{minipage}{0.75\hsize}
  \centering
  \includegraphics[width=\columnwidth]{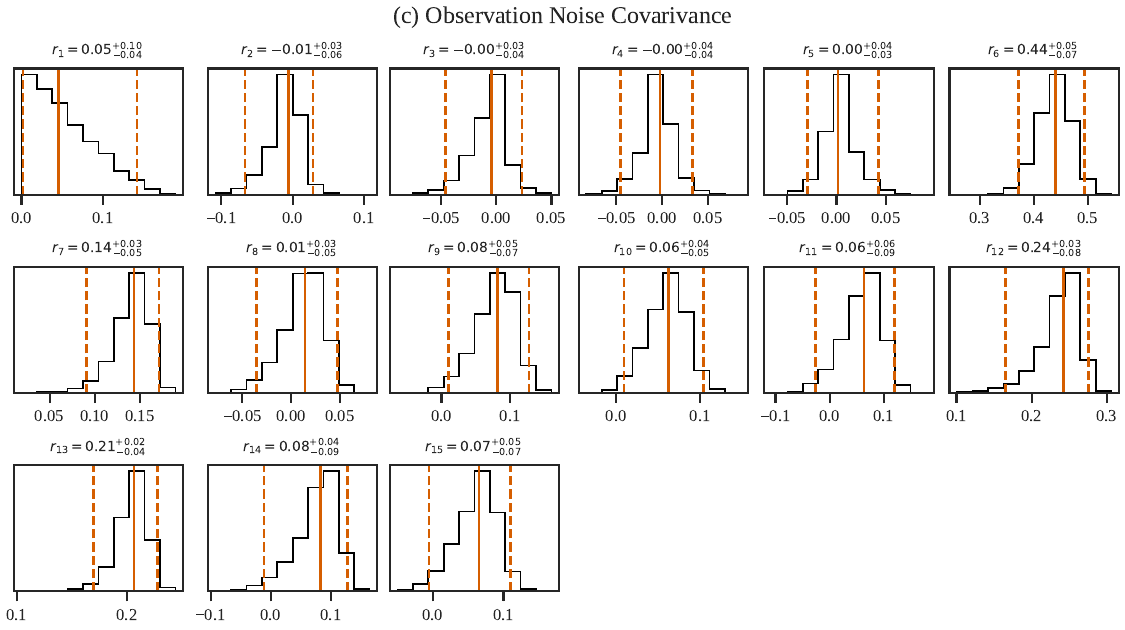}
 \end{minipage}
 \end{center}
 \caption{
  Posterior probability distributions of the linear state-space model parameters estimated via the MCMC sampling.
  Rows (a), (b), and (c) display the distributions for the VAR transition coefficients, the observation matrix elements, and the observation noise covariance matrix elements, respectively.
  The layout within each section corresponds to the matrix structures defined in equations~\eqref{eq:lgssm_var_coefficients}, \eqref{eq:j1820_lgssm_loading_matrix}, and \eqref{eq:j1820_lgssm_observation_noise_covariance}.
  The numbers printed above each sub-panel indicate the estimated median along with the 2.5\% and 97.5\% quantiles.
 }
 \label{fig:mcmc_sample_distributions}
\end{figure*}

Figure~\ref{fig:mcmc_sample_distributions} shows the posterior distribution of the
Bayesian inference conducted in \S~\ref{section:model_configuration}. For each sample,
using the K\'{a}lman filter smoothing, we can construct the light curves of the
observation variables and compare them to the observed data
(figure~\ref{fig:j1820_lgssm_lightcurve_observation}). The smoothed observation
variables describe the data well in all bands. In addition, we can also construct the
light curves of the state variables, or the physical components
(figure~\ref{fig:j1820_lgssm_lightcurve_smoothed_state}).  This demonstrates the
advantage of using the observation model in the state-space model; no such light curve
can be obtained through classical approaches.

We used the normal distribution as the prior distribution of the VAR matrix
elements. This ignores the stationary conditions to be met, but we expected 
that the MCMC process would reject the non-stationary samples. Indeed, in the 
posterior distribution (figure~\ref{fig:mcmc_sample_distributions}), most of 
the coefficients are sampled within the range of $-1$ to $+1$, which implies 
the model stationarity is likely satisfied as a result.

\bibliographystyle{aasjournal}
\bibliography{main}

\end{document}